\documentclass[acmsmall]{acmart}

\usepackage{booktabs, siunitx, hyphenat} 
\usepackage[raggedrightboxes]{ragged2e}

\setcopyright{rightsretained}
\acmJournal{PACMHCI}
\acmYear{2025} \acmVolume{9} \acmNumber{2} \acmArticle{CSCW110} \acmMonth{4} \acmPrice{}\acmDOI{10.1145/3711008}

\begin{document}

\title{Identifying the Barriers to Human-Centered Design in the Workplace: Perspectives from UX Professionals}


\author{Tim Gorichanaz}
\email{gorichanaz@drexel.edu}
\orcid{0000-0003-0226-7799}
\affiliation{%
  \institution{Drexel University}
  \streetaddress{3675 Market Street}
  \city{Philadelphia}
  \state{Pennsylvania}
  \country{USA}
  \postcode{19104}
}

\begin{abstract}
Human-centered design, a theoretical ideal, is sometimes compromised in industry practice. Technology firms juggle competing priorities, such as adopting new technologies and generating shareholder returns, which may conflict with human-centered design values. This study sought to identify the types of workplace situations that present barriers for human-centered design, going beyond the views and behaviors of individual professionals. Q methodology was used to analyze the experiences of 14 UX professionals based in the United States. Five factors were identified, representing workplace situations in which human-centered design is inhibited, despite the involvement of UX professionals: Single-Minded Arrogance, Competing Visions, Moving Fast and Breaking Things, Pragmatically Getting By, and Sidestepping Responsibility. Underpinning these five factors are the dimensions of speed and clarity of vision. This paper demonstrates connections between the literature on UX ethics and human-centered design practice, and its findings point toward opportunities for education and intervention to better enable human-centered and ethical design in practice.
\end{abstract}




\maketitle

\section{Introduction}

Since its inception in the 1980s \citep{GouldLewis1985}, human-centered design has become the preferred paradigm for designing interactive products \citep{Sharp2023} and increasingly other sorts of products and services \citep{Han2022}. In theory, that is, perhaps more than in practice. 

Consider how throughout 2023, many companies and startups scrambled to infuse their current offerings with generative AI and develop new AI-based products. In January 2023, for example, Microsoft announced plans to add AI capabilities into every one of its products \citep{Sinclair2023}, and soon Google followed suit \citep{Love2023}. In the first half of 2023, generative AI startups collectively raised \$14.1 billion in disclosed equity funding, more than in all of 2022 and despite an overall contraction in investing \citep{CBI2023}. These are examples not of human-centered design but of technology-driven design. 

To be sure, there are high-profile examples of successful industry products developed with human-centered design. The Nest smart thermostat, for instance, was created in 2010 with a human-centered approach in which every aspect of the device's user experience was prototyped and iterated upon \citep{Fadell2022}; the thermostat was a major commercial success and is credited with influencing later smarthome products \cite{Mylavarapu2016}. Yet academic research on user experience (UX) practice continually shows that there are barriers and challenges to doing human-centered design in industry \citep[e.g.,][]{Feng2023,Nielsen2023,Poiroux2023,Watkins2020}. 

If human-centered design is perceived as the gold standard for designing usable digital products and services, why is it not always successful---or even used at all---in practice? Where does it fall short? Answering that question requires practice-led research \citep{Kuutti2014} and can serve to further improve theorization in HCI \citep{Gray2014}. Because UX is a major realm of practice for HCI, this paper follows the precedent of recent literature in focusing on the workplace practices of UX professionals \citep[e.g.,][]{Chivukula2020,Feng2023,GrayChiv2019}.

This paper sheds light on why human-centered design sometimes fails in practice---why a non--user-centered design process is sometimes followed despite the involvement of UX professionals. It presents a study using Q methodology, a mixed-methods approach to studying subjectivity, to analyze the experiences of 14 U.S.-based UX professionals. The study responds to the research question: \emph{In the experience of UX professionals, what workplace contexts inhibit human-centered design from taking place?} The resulting analysis suggests five workplace contexts that interfere with human-centered design: Single-Minded Arrogance, Competing Visions, Moving Fast and Breaking Things, Pragmatically Getting By, and Sidestepping Responsibility. Underpinning these five factors are two dimensions: speed and clarity of vision. While the results from this study primarily show how human-centered design goes wrong, the framework generated from these results offer suggestions for how to make it go right: by promoting a design culture within an organization and by cultivating organizational humility---the practices of unlearning and rethinking. 

This paper makes two primary contributions. First, it outlines five organizational contexts that lead away from human-centered design. This categorization is valuable for education, awareness and diagnostics. While these contexts are thematically resonant with the prior literature, this framework is novel, as is the methodology used to arrive at it. Second and furthermore, this paper provides a higher-level framework with the dimensions of speed and clarity of vision to describe when human-centered design fails and when it can best succeed. This can be used as a foundation for further research on UX practice in HCI, as well as a guide for cultural change within organizations. This work may helpfully direct the efforts of UX designers and other stakeholders---as well as educators striving to prepare their students for the realities of the workplace.


\section{Related Work}

To frame this study, this section reviews the literature on the human-centered design paradigm, on UX professional work in practice, and on ethical issues in UX practice. 

\subsection{The Paradigm of Human-Centered Design}

When computers were first created, there wasn't much attention given to usability. At that time, computers were only used by experts---often the very people who built the machines. But by the 1960s, the roles of maker, programmer and user had diverged, and soon the field of HCI was born.

For at least the first decade, HCI was narrowly focused on usability engineering---making systems more effective, efficient, safe, learnable and so on \citep{Grudin1990}. But soon scholars began rallying for a wider scope, which set the groundwork for the paradigm of human-centered design to emerge. The seminal paper for human-centered design was the 1985 ``Designing for Usability'' by Gould and Lewis, in which the authors proposed three principles to improve usability: an early focus on users and tasks, empirical measurement, and iterative improvement \citep{GouldLewis1985}. Soon after, Don Norman published \emph{The Psychology of Everyday Things}, later revised as \emph{The Design of Everyday Things}, in which he summarized the key principles of human-centered design: addressing the root problems rather than symptoms, focusing on human needs and capabilities, thinking in systems, and continually testing and refining, or iterating upon, solutions \citep{Norman2013}. More generally, human-centered design provides a process and methods to respond to human needs and contexts, resulting in more efficient, productive and positive experiences \citep{Sharp2023}. 

Human-centered design undoubtedly enables usability and high-quality user experiences, but in recent years it has become clear that the methods of human-centered design is not enough to reliably lead to products that meet human needs, particularly when conceptualizing ``needs'' in a deep, moral sense. 

First, recall that though human-centered design is considered an ideal in UX education, research and professional philosophy, it is not always carried out in practice, as discussed in the introduction. Next, consider that even when human-centered design is followed, it is not always successful. Early critics observed that human-centered design methods may be superficially or incorrectly implemented, such as by defining narrowly-construed tasks rather than contextualized human activities \citep{Gasson2003,Norman2005,Bannon2011}. More recently, scholars have noted additional issues with the paradigm: human-centered design creates products for the market, thus serving short-term desires rather than long-term human futures \citep{chapman2021,DunneRaby2013}; in human-centered design, innovation itself is considered progress, disincentivizing systemic change \citep{Harris2021,Pierce2021}; and human-centered design is anthropocentric, ignoring the other species we share our world with as well as the needs of our planet itself \citep{Light2017,Wakkary2021}.

Working to overcome these issues gave rise to the ``third wave'' of HCI, which has informed the philosophy and practice of human-centered design \citep{Bodker2006,Harrison2011,Sharp2023}. Over the past two decades, HCI has become more concerned with the contextual, social, affective and emergent dimensions of interaction \citep{Bodker2006,Gaver2022,Harrison2011}. Moreover, deeper questions such as those of worth, identity and meaning have come to the fore \citep{Cockton2006,Cockton2020,Fallman2011,Kaptelinin2018,Mekler2019}. Frameworks such as participatory design and value sensitive design have matured, and naturalistic research became more common \citep{Rogers2017}. 

Most recently, a new formulation of human-centered design has been proposed accounting for all this under the banner of ``humanity-centered design.'' In his 2023 book \emph{Design for a Better World} \citep{Norman2023}, Don Norman articulates five tenets of humanity-centered design, which he describes as compatible with but going beyond those of human-centered design: addressing the root problem, focusing on the ecosystem, taking a long-term systems perspective, continually testing and refining, and designing with the community as much as possible. However, this and similar descriptions of humanity-centered design suffer from internal contradictions that must be resolved for further progress to be made \cite{Gorichanaz2024altchi}.

\subsection{UX in Practice}

HCI has shifted over the past few decades from largely ahistorical, lab-based research emphasizing technology, to contextualized, in-the-wild research emphasizing the relationship between technology and people. In part, this shift is encapsulated in the emergence of CSCW as a research field. As part of this shift, HCI has become increasingly concerned with the relationship between theory and practice \citep{Kuutti2014}. Authors such as \citet{Gray2014} have observed this ``turn to practice'' in HCI, calling for researchers and professionals to work together to unlock the latent synergy between theory and practice in HCI. 

Perhaps the principal site of practice in HCI is the field of user experience (UX), which concerned with designing interactive systems for deployment, typically in a commercial context. Human-centered design lies at the heart of the ethos of the UX profession \citep{Nielsen2017}.

Though UX activities date back to 1950, the term ``user experience'' was not coined until 1993, and it wasn't until the new millennium that UX activities were strongly formalized and professionalized \citep{Nielsen2017}. As one example in this regard, the Usability Professionals Association was founded in 1991 and renamed to the User Experience Professionals Association in 2012. Consonantly, the number of UX professionals (by whatever name) has also grown tremendously over the decades. \citet{Nielsen2017} estimates that there were about 10 UX professionals globally in 1950, about 1,000 in 1980, 100,000 in 2010, and 1 million in 2020; and Nielsen forecasts that by 2050 there will be 100 million UX professionals. Already, UX has been adopted in many organizations worldwide, and UX is increasingly a household term (interestingly and perhaps unfortunately, far more so than HCI). 

UX activities may take place in a department within a corporation (``in-house'') or at a design firm doing UX work on a contract basis for external clients (``agency''). In both settings, UX work falls into a number of subspecialties; the major bifurcation in the profession is between UX research and UX design, though the growth of UX has necessitated additional roles such as UX management and UX strategy \citep{Rosala2020}. That said, the job of a given UX professional may cross these subspecialties depending on their organizational context. For example, a large software company may employ UX researchers as such, while a small design agency may only employ one or two people responsible for all aspects of UX. Reflecting this diversity, Nielsen Norman Group reports that UX professionals hold over 200 different job titles \citep{Rosala2020}. 

According to data from 2019, UX professionals typically have an educational background in design, and many have skills in front-end coding. ``Soft skills'' such as negotiation, management, teamwork and communication are increasingly vital for getting and succeeding in a UX job \citep{Rosala2020}. Increasingly, UX professionals are educated on HCI theory as well as concepts, tools and methods specific to UX. Still, UX professionals generally do not consider themselves as belonging to a professional community outside their employer \citep{Inal2020}. 

Given the complexity of software development, today UX work involves more collaboration than it has in the past. UX professionals routinely collaborate with other UX professionals and management \citep{Kuang2023,Shukla2024}, as well as with software engineers who do not have UX expertise \citep{Frishberg2020,Yang2018,Yang2020}, both of which introduce challenges. Tools such as Figma (launched in 2016) have emerged to facilitate such collaboration, but the ``handoff'' of designed assets to developers remains a challenge \citep{Maudet2017,Yang2020,Feng2023}, and new challenges have emerged such as the management of design systems \citep{Feng2023} and designing with and for AI \citep{Cho2024,Feng2023b,Feng2023c,Varanasi2023}.

Related to challenges with collaboration, prior research has shown how UX work may indeed be thwarted unintentionally by dynamics within an organization or by a client, further evincing the disconnect between theory and practice in HCI/UX and shedding some light on how ``non--human-centered'' designs emerge. \citet{Gray2015} found that companies may grow their UX capacity while other areas of the company remain effectively hostile to UX. For example, client or executive priorities such as efficiency and profit may be at odds with UX professional values and goals regarding usability and accessibility \citep{Gray2015,Patel2020}. Research by \citet{Poiroux2023} shows how UX designers may indeed intervene in product development at various levels (technical implementation, interface, product and organizational), but ultimately tend to defer to client wishes, which may go against UX professionals' recommendations. 

Responding to this issue, \citet{Frishberg2020} discuss the importance of UX collaboration with non-UX parts of the organization. Some of the barriers for doing so include: achieving appropriate confidence levels, trust in UX, commitment and support structures, trade-offs in UX, handoff practices, and UX management \citep{Nielsen2023}. Another aspect of this issue, discussed by \citet{Watkins2020}, is that a designers' philosophy, as inculcated in their educational experience, may be impossible to enact in real-life organizational settings. Design education, then, should rather help students form a bridge between their emerging design philosophy and the exigencies of practice. 

\subsection{UX Practice and Ethics}

The UX profession has grown tremendously over the past decade but has yet to be entirely integrated within the technology industry. For instance, the values of human-centered design may conflict with other interests, such as growth and profit. The industry does not yet have a satisfactory way to navigate such tensions, which amount to ethical tensions. \citet{Poiroux2023} ``perceived a strong ethical discourse from designers but that discourse seldom translated into concrete interventions.'' As other researchers show as well \citep{Gray2015,Nielsen2023,Watkins2020}, ethics is a site of the theory--practice disconnect par excellence. Though UX professionals are increasingly educated on ethics \citep{Fiesler2020,Findeli2001,sonneveld2016,Vilaza2022}, that education has yet to take root in the ``real world'' \citep{Gray2021,Lindberg2020,Nelissen2022,Pillai2022,Varanasi2023,Wong2023}. 

While human-centered design and design ethics are not coterminous, there is substantial overlap between working to create products that meet human needs (particularly when conceptualized through the lens of third-wave HCI\footnote{The emergence of third-wave HCI and recent discussions of humanity-centered design, inter alia, also suggest an increasing alignment of human-centered design and design ethics.}) and working to create products that promote human flourishing, relationships and the good society (which are at least some of the goals of ethics). Indeed, many theorists have argued that design is an inherently moral enterprise \citep{Findeli1994,Flusser1999,Tonkinwise2004,Verbeek2006,Winner1980}, thus concluding that ``design is either ethics materialized or ethics neglected'' \citep{Fry2003}. And while human-centered design may not in itself be sufficient to ensure ethical design outcomes, human-centered design and design ethics are of a kindred spirit. 

\citet{NunesVilaza2022} present a literature review of ethics in HCI research from 2010 to 2020, identifying three major areas of implications: research ethics, ethical choices in interface design, and questions of responsibility. In the literature, even though ethical best practices and theories are emerging, such as increasing user autonomy and knowledge and considering how to support user well-being and safety, tensions remain regarding who among the various stakeholders is responsible for the outcomes of design and product development. So far, most of the papers in this area center their discussions on individuals, whether designers or the public, though a few look at the roles of institutions in sociotechnical morality. Because individuals are limited in what they can do given the complexities of the technology industry, \citet{NunesVilaza2022} suggest that a way forward ``could be to establish more robust and evidence-based policies and redistribute responsibility across other entities,''  calling for ``an overarching ethical framework harmonizing institutional, collective and individual roles.''

A similar argument has been made by \citet{Arledge2019} in the trade literature. Arledge observes that designers have little influence over the business and infrastructure, though these are inextricable from issues in design ethics. He suggests that intervening at the level of infrastructure would have the most influence over outcomes, and so he calls for designers to work to influence national policy. 

\subsubsection{Ethically Mediating Factors in UX Practice}

Looking more granularly at research on the moral complexity of UX, one influential model has been the account of ethical mediation in UX practice by \citet{GrayChiv2019}. This account models three ethically mediating factors in design outcomes: individual practices (the designer's own character), organizational practices (the structure and purpose of the organization), and applied ethics (knowledge of ethics, professional codes of ethics, etc.). Ideally these three factors would be synergistic toward ethical outcomes, but in reality they tend to be in tension. 

Other research on ethics in UX practice can more or less be mapped to these mediating factors. Regarding individual practice, researchers have: examined the identity claims and beliefs that influence whether and how technology practitioners make moral decisions \citep{Chivukula2021b}; profiled the subtle ways in which UX professionals may practice ``soft resistance,'' acting against their employers to protect customer interests \citep{Wong2021}; and documented how UX practitioners promote AI ethics within their organizations \citep{Morley2023,Wang2023}.

Regarding organizational practice, researchers have delineated some of the complexities that influence designers' practical agency in their roles, such as the role of UX in the enterprise and how conflicts are negotiated in decision making. This work finds a lack of support for ethical design activities across the organization \citep{Chivukula2020,SanchezChamorro2023}. Moreover, existing structures such as toolkits may not adequately address the ethical issues they intend to address \citep{Deng2022,Wong2023}. Individuals may seek to promote ethics unilaterally, but this comes with risk and sometimes personal cost \citep{Ali2023,SanchezChamorro2023}. Overall, this work demonstrates that UX professionals value the user over other stakeholders, yet this is at odds with the actual power that UX has within an organization \cite{Ali2023,Chivukula2020,SanchezChamorro2023,Wang2023,Wong2021,Yildirim2023}.

Regarding applied ethics, researchers have sought to contribute to human knowledge about the nature and origins of unethical designs. For example, \citet{Watkins2020} profile the personal philosophies inculcated in designers through their education---which by and large have an ethical and human-centered emphasis---but show that these philosophies are generally suppressed in practice because of the dynamics discussed above. Work in this area includes that analyzing dark patterns \citep{Gray2018,Gray2023,Nelissen2022}, which are deceptive, manipulative or coercive tactics that put business interests above user interests. Similar to that line of work, \citet{Gray2020} articulates the properties of products that users deem ``asshole design,'' which overlaps with but is distinct from designs with dark patterns. \citet{Varanasi2023} discuss how the values of Responsible AI are conceptualized and operationalized in tech firms.

\subsubsection{Drilling Down on ``Applied Ethics'': Macroethics and Microethics}

\citet{GrayChiv2019} describe applied ethics as the knowledge and codification of ethical principles, but this may be only part of the fuller picture of applied ethics. Recent work in engineering ethics and medical ethics has identified two frames of reference for applied ethics: macroethics and microethics. Macroethics describes ethical issues in a field, organization, profession, society, etc., in line with how \citet{GrayChiv2019} describe applied ethics \emph{tout court}. On the other hand, microethics describes ethical issues internal to professional practice \citep{Herkert2001}. 

Historically, discussions of applied ethics (including in UX) have tended to focus on macroethics. These discussions have been broad in scope and conceptual in orientation, removed from the processes they have sought to describe, influence and evaluate---and thus difficult to apply in real-life situations \citep{Bezuidenhout2021}. Thus \citet{Bezuidenhout2021} argue for technology professionals to turn toward microethics, as has been done in the medical field, as a strategy for embedding ethics into professional practice. Microethics is about seeing the ethical relevance of even small acts within a structure, such as \citet{Wong2021} observes in terms of soft resistance. 

Returning to the framework of ethical mediators proposed by \citet{GrayChiv2019}, it is evident that their notion of ``applied ethics'' is indeed exclusively one of macroethics, in line with the broader trend in the literature. But the frame of microethics would seem to bridge the three mediators. As such, a concerted focus on microethics in HCI research may illuminate new paths toward better resolving the tensions that arise in UX practice.

In HCI, one extant proposal that exemplifies microethics includes ``in-action ethics'' \citep{Frauenberger2017}, a situated and responsive framework for ethics. In practice, in-action ethics involves ongoing reflection both individually and as a design team, as well as reflective discussions (e.g., discussing whether $x$ was the right thing to do, what might have happened if we did $y$, what ``good'' means in this situation, etc.). 

One limitation of the in-action ethics framework is that it is primarily rooted in HCI design research and may not account well for the vicissitudes of UX professional practice. Still, the themes of in-action ethics are resonant with the findings from interview research showing that UX professionals do not use formal methods or tools for ethics but rather rely on ad-hoc, dialogue-centric ways of engaging with ethical reasoning, navigating competing interests, discussing ethical concepts, etc. \citep{dindler2022,Varanasi2023}, also echoing earlier findings \citep{Chivukula2020}. 

\section{Research Approach}

The prior work in human-centered design, UX practice and UX ethics demonstrates that UX practice is complex and social---the domain of trade-offs and applied ethics. This body of research points to the need to develop the philosophy of human-centered design and the ethics of design in more realistic, practice-led ways, as well as the need to continue to improve education as the UX field professionalizes and diversifies to ensure a match between educational components and the realities of the workplace. That being the case, there are opportunities for research to further examine the complexities of UX workplace situations through practice-led research. One promising framing for such work is microethics, which looks at ethical issues as they arise in lived situations in a particular context. Examining how human-centered design is operationalized (or not) in real-world settings is one opportunity for research within the framing of microethics.

This study used Q methodology to address the research question: \emph{In the experience of UX professionals, what workplace contexts inhibit human-centered design from taking place?} Addressing this question sheds light on microethical issues in UX practice, including how non--human-centered products and features may emerge from teams that include UX professionals. This study builds upon prior work on ethics in UX practice by characterizing work contexts and situations, going beyond characterizing professionals' own activities and identities.

Q methodology, first developed in the 1930s, provides a rigorous set of methods for the systematic, exploratory study of subjective experiences. It uses both quantitative and qualitative tools to identify a range of viewpoints on a complex issue and reveal areas of consensus and disagreement. The results of Q-methodological studies include the factors that reflect clusters of participants' experiences \citep{McKeown2013}. As such, Q methodology reveals a range of shared understandings across an issue, which together offer ``a comprehensive snapshot of the major viewpoints being expressed by [the] participant group'' \citep[p.~85]{Watts2005}.

In HCI, Q methodology has previously been used to describe attitudes toward autonomous public transit \citep{Detjen2021} and show how people with chronic illnesses use technology to manage their health \citep{OLeary2015}. Though there are not many examples of the methodology in the HCI literature to date, two additional articles argue for its usefulness (to say nothing of such articles in other fields, chiefly psychology, where the methodology originated). Most recently, \citet{Allison2022} has discussed how Q methodology can fruitfully identify differing perspectives on how people think about technology, such as in computing education research; and earlier, a modified version of Q methodology, dubbed HCI-Q, was proposed as a design research method \citep{OLeary2013}. This study uses the standard approach to Q methodology as developed in psychology and as appropriate to a study aimed at understanding a group of people's subjectivities across a complex issue. 

Q methodology involves a series of sequential steps \citep{McKeown2013}, some of which introduce terms of art for the methodology:

\begin{enumerate}
	\item framing a research question
	\item developing the set of statements to be sorted by participants (the \emph{Q-set})
	\item selecting the participants (the \emph{P-set})
	\item asking each participant to sort the statements along a continuum from ``most disagree'' to ``most agree'' (the \emph{Q-sort})
	\item conducting a factor analysis of the Q-sorts (numerical values are assigned to levels of agreement in order to conduct this analysis; e.g., --1 for somewhat disagree, 0 for neutral, +3 for very strongly agree, etc.)
\end{enumerate}

The following subsections describe how the Q-set for this study was developed, how the P-set was selected and Q-sorts gathered, and how the Q-sorts were analyzed.

\subsection{Developing the Q-set}

Once the research question is established, the next step in Q methodology is to develop the Q-set, or the set of statements that participants will sort. This begins with identifying the so-called \emph{concourse}, or the complete set of prospective statements from which the Q-set is a sample. The concourse is a hypothetical construct that represents the breadth of possible discourse on a topic. The Q-set is a sample from the concourse, generally comprising 40--80 statements, and it may be sampled in a structured or unstructured way \citep[pp.~23--24]{McKeown2013}. The Q-set should be broadly representative of the range of possible views in the concourse. 

In this study, the concourse included the breadth of possible situations and contexts that may contribute to a product or feature being less human-centered (and as a corollary, less ethical) than it might otherwise be. Put differently, the concourse was everything that might go wrong throughout the process of human-centered design.

I used a semi-structured, iterative approach to develop the Q-set for this study, gathering prospective statements from a variety of sources. I drew on my five years of experience teaching human-centered design at the undergraduate and graduate levels, as well as textual sources. These sources included the textbooks \emph{Interaction Design} \citep{Sharp2023} and \emph{Introduction to Information Systems} \citep{Wallace2017}, literature describing various challenges in human-centered design \citep[e.g.,~][]{Cornet2020,Gulliksen1999}, and the literature describing practical and ethical issues in UX surveyed in the ``Related Work'' section above. 

I began by listing potential statements in no particular order over a period of several weeks in Spring 2023. Once I identified two dozen or so, I sorted them according to the stages of the design process and software development life cycle that they related to. These categories included: research, design, development, management, ethics, and user behavior. Once the statements were categorized in this way, I generated more statements so that each category had several statements. Again over a period of weeks, I reconsidered the list of statements, combining similar ones and rewording others for clarity. I ran a small pilot study with colleagues who had industry UX experience to test the Q-set and made further modifications thereafter.

The final Q-set of 39 statements is shown in Table~\ref{table:statements}, along with the factor loadings (described below in the Findings).

\begin{table}[]
	\caption{This table shows the 39 statements that comprised the Q-set, along with the resulting factor (F) loadings. In this table, +3 indicates that the participants in that factor on average most agreed with that statement (rated it as ``Most likely to lead to non--human-centered design''), while --3 indicates that they disagreed most with that statement (``Least likely to lead to non--human-centered design''). Distinguishing statements (* p < 0.05, ** p < 0.01) reflect the statements that a factor group ranked significantly different from another factor group.}
	\label{table:statements}
	\resizebox{\textwidth}{!}{%
\begin{tabular}{@{} l l *5{S[table-format=2.0, table-space-text-post ={***}]} @{}}
	\toprule 
\# & Statement                                                                    & F1   & F2   & F3   & F4  & F5   \\
	\midrule
1  & User needs conflict with each other                                          & -1   & 0    & 0    & 2   & -1   \\
2  & Not doing research to understand user needs and values                       & 2    & 3    & 3    & 2   & 2    \\
3  & Missing an important stakeholder group during research                       & 1    & 0    & 2    & -2  & -1   \\
4  & Too much data collected during research                                      & -1   & -3   & -3   & -2  & -3   \\
5  & Disagreements interpreting research insights/results                         & 1    & -2** & 2    & 0   & 1    \\
6  & Planning to implement features later that are better aligned with user needs & 0    & -1   & 2*   & 0   & -2*  \\
7  & Ignoring insights from user research                                         & 2    & 1    & 2    & 2   & 2    \\
8  & Not enough different ideas explored early in the design process              & 1    & -2   & -1   & -1  & 1    \\
9  & Lack of iteration                                                            & 2    & 0    & -2** & 2   & 1    \\
10 & Not enough time/resources spent in prototyping                               & 0    & -1   & 0    & 0   & 1    \\
11 & Not enough time/resources spent in testing                                   & 1    & -1   & 2    & 0   & 2    \\
12 & Some requirements had to be frozen and could not change with iteration       & -1   & -1   & -1   & 2   & 0    \\
13 & Compromises during technical implementation                                  & -2*  & 0    & -1   & 2   & 0    \\
14 & Mistakes during technical implementation                                     & -1   & 0    & 0    & 0   & 0    \\
15 & Goals or values changed during the project                                   & 1    & 0    & -1   & 0   & -2   \\
16 & Lack of stakeholder involvement at the necessary times                       & 1    & 2    & -2   & 0   & -1   \\
17 & Scope creep                                                                  & -2   & -2   & 2    & 0   & 0    \\
18 & Prioritizing business goals over user needs                                  & 2    & 1    & 1    & 1   & 1    \\
19 & Fulfilling a non-negotiable client request                                   & -1   & 0    & 1    & -1  & 0    \\
20 & Stakeholders not seeing the value of user-centered design                    & 3    & 2    & 1    & 1   & -1** \\
21 & Miscommunication among team members                                          & 0    & 0    & -1   & 0   & -1   \\
22 & Sabotage, vendetta or retribution                                            & 0*   & 2    & -2*  & -3* & 2    \\
23 & Lack of the necessary people taking responsibility                           & 0    & 1    & 0    & 1   & 2    \\
24 & Politics between teams involved                                              & -1   & 2    & 0    & 1   & 1    \\
25 & Certain requirements were non-negotiable                                     & -2** & 1    & 1    & -1  & -1   \\
26 & Lack of time                                                                 & 0    & 1    & 1    & 1   & -2** \\
27 & Lack of budget                                                               & -2   & 1    & 0    & 1   & -2   \\
28 & Resistance to change                                                         & 2    & 1    & -2** & 1   & 1    \\
29 & Not trying to create a user-centered product                                 & 1    & 2    & -1** & 3   & 3    \\
30 & Personnel changes during project                                             & -1   & 0    & -2   & -1  & -2   \\
31 & Lack of commitment or interest among team members                            & 0    & 1    & 1    & -1  & 1    \\
32 & Using ethical toolkits incorrectly                                           & 0    & -2** & 0    & -1  & 2*   \\
33 & Adhering to regulations/laws                                                 & -2   & -2   & -2   & -2  & 0**  \\
34 & Lack of exploring ethical values                                             & 2    & 2    & 1    & -2* & 0    \\
35 & Lack of ethical knowledge                                                    & 1    & -1   & 1    & -2  & 0    \\
36 & After launch, product/feature was used in unpredictable ways                 & -3** & -1   & 0    & -1  & 0    \\
37 & The product evolved over time                                                & -1   & -1   & -1   & -2  & -2   \\
38 & Something changed in the outside world                                       & -2   & -2   & -1   & 1*  & -1   \\
39 & Problems only appeared after scaling up                                      & 0    & -1   & 0    & -1  & -1  \\
	\bottomrule
\end{tabular}}
\end{table}

\subsection{Selecting the P-set}


In Q methodology, participants (collectively called the P-set) are recruited to sort the Q-set (in an activity called the Q-sort). Small P-sets are typical in Q methodology; indeed, single-case studies can be particularly illuminating for certain research questions \citep{McKeown2013}. Methodological guidance suggests that the P-set should be smaller than the Q-set; typically, a ratio of 3-to-1 statements to participants is used, and thus many Q studies involve 12--20 participants \citep{Webler2009}. (That guidance would suggest a P-set size of 13 for the present study, whose Q-set includes 39 statements.) However, \citet{Watts2005} suggest that the size of the P-set is less important than its strategic sampling. 

In this study, the P-set was assembled through purposive sampling of UX professionals with industry experience based in the United States. Within these parameters, I sought to recruit a P-set that was diverse in terms of experience level, role and work context. Still, recruitment was restricted to professionals working in the United States to minimize (so much as possible) other cultural factors as confounding variables. 

I recruited participants primarily through LinkedIn to identify second- and third-degree connections, and secondarily through finding public postings of UX profiles and portfolios. I sent emails to prospective participants, inviting them to take part in the anonymous online Q-sort. Participants could choose to optionally to be entered in a lottery to win an Amazon gift card upon completing the Q-sort. The expected value for completing the survey was \$5. 

All in all, I sent 180 invitations, and these resulted in 14 valid responses, equating to an 8\% response rate. The P-set successfully represents a range of types of UX professionals in terms of experience level, role and work context, as shown in Table~\ref{table:participants}.

\begin{table}[]
	\caption{Participants in the study (P-set)}
	\label{table:participants}
\begin{tabular}{@{}lllll@{}}
\toprule
Participant & UX Experience      & UX Role(s)                 & Context          & Business Type \\ \midrule
P1          & 2–5 years          & Strategy, Writing          & Freelance        & B2B,B2C       \\
P2          & 10+ years          & Research                   & Freelance        & B2B,B2C       \\
P3          & \textless{}2 years & Design, Strategy, Product  & Freelance        & B2B           \\
P4          & 2–5 years          & Design, Research, Strategy & In-house         & B2B,B2C       \\
P5          & 10+ years          & Design                     & In-house         & B2C           \\
P6          & 10+ years          & Design, Research, Strategy & In-house         & B2B,B2C       \\
P7          & \textless{}2 years & Design                     & Freelance        & B2C           \\
P8          & 10+ years          & Design, Strategy           & In-house         & B2C           \\
P9          & \textless{}2 years & Research                   & In-house         & B2C           \\
P10         & 5–7 years          & Design                     & In-house         & B2C           \\
P11         & 7–10 years         & Engineering                & In-house         & B2B,B2C       \\
P12         & 2–5 years          & Design                     & Agency           & B2B,B2C       \\
P13         & \textless{}2 years & Design                     & In-house         & B2B,B2C       \\
P14         & 2–5 years          & Design, Research           & Agency, In-house & B2C           \\ \bottomrule
\end{tabular}
\end{table}

\subsection{Gathering Q-sorts}

This study was implemented as an anonymous online survey using Qualtrics. Data was collected in July and August 2023. Participants spent 7--31 minutes completing the survey (mean 19 minutes).

The Q-sort itself was a ``Pick, Sort and Rank'' question, in which the Q-set statements appeared on individual cards that participants dragged into one of five boxes. The prompt question for the Q-sort was, ``Based on your experience, why might a product team end up launching a product or feature that goes against user needs?'' The boxes that participants chose from were labeled: 

\begin{itemize}
	\item Very likely to lead to non--human-centered design (+2)
	\item Likely to lead to non--human-centered design (+1)
	\item Neutral (0)
	\item Unlikely to lead to non--human-centered design (-1)
	\item Very unlikely to lead to non--human-centered design (-2)
\end{itemize}

Participants were also asked to identify the statements they felt were most likely (+3) and most unlikely (-3) to lead to a non--human-centered design, thereby showing the extremes in their subjectivity with this Q-set. 

In this study, Q-sorts were not constrained to a quasi-normal distribution (as is sometimes done in Q-methodological research), as e.g. \citet{Brown1971} has shown that doing so is unnecessary.

After the Q-sort, participants were invited to share a story of a time when they or a team they knew of created a product or feature that was not as user-aligned as it could have been, along with what led to that issue. (Participants were also reminded here to keep the story anonymous or pseudonymous.) These stories served as an opportunity for participants to reflect further and share more about their sort and their viewpoints. Ten participants shared such a story, each one 85 words on average. Finally, certain demographic information was collected: years of experience in UX, primary UX role type, UX setting, and types of clients (see Table~\ref{table:participants}). 

\subsection{Analyzing the Q-sorts}

In Q methodology, statistical factor analysis is used to explore correlations across the Q-sorts and discern factors that represent clusters of experiences. While methodological texts caution relying on general guidelines for extracting factors and suggest using judgment instead, best practices for Q methodology suggest that factors should have an eigenvalue greater than 1 and that each factor should have at least two defining Q-sorts (that is, Q-sorts which load more strongly on that factor than on other factors) \citep{McKeown2013}. 

To manage the statistical analysis of the Q-sorts, I used the software package Ken-Q Analysis Desktop Edition (KADE v1.2.1) for macOS. I used principal components analysis, which showed that five factors had an eigenvalue greater than 1. I applied varimax rotation to arrive at a terminal solution, which is summarized in Table~\ref{table:factorchars}. All but one of the factors had at least two defining Q-sorts; Factor~4 had one defining Q-sort. 

\begin{table}[]
	\caption{Characteristics of factors (F) after rotation}
	\label{table:factorchars}
\begin{tabular}{@{}lllllll@{}}
	\toprule
Characteristic                       &        & F1     & F2     & F3     & F4     & F5 \\
	\midrule
Number of defining Q-sorts 		     &        & 3      & 3      & 2      & 1      & 2     \\
Average rel. coefficient             &        & 0.8    & 0.8    & 0.8    & 0.8    & 0.8   \\
Composite reliability                &        & 0.923  & 0.923  & 0.889  & 0.8    & 0.889 \\
S.E. of factor Z-scores              &        & 0.277  & 0.277  & 0.333  & 0.447  & 0.333 \\
\% Explained variance (EV)           &        & 20     & 19     & 12     & 11     & 13    \\
Cumulative \% EV                     &        & 20     & 39     & 51     & 62     & 75    \\
	\midrule
Factor~score correlations            & F2     & 0.3322 &        &        &        &    \\
                                     & F3     & 0.2295 & 0.2668 &        &        &    \\
                                     & F4     & 0.108  & 0.351  & 0.3201 &        &    \\
                                     & F5     & 0.3877 & 0.1955 & 0.2232 & 0.1469 &   \\
	\bottomrule
\end{tabular}
\end{table}

One result of the factor analysis in Q methodology is a set of ``composite Q-sorts,'' or hypothetical Q-sorts that best express each factor (see Table~\ref{table:statements}). Along with the defining Q-sorts themselves, as well as the defining statements for each factor, these composite Q-sorts served as the grounds from which I interpreted the meaning and characteristics of each factor. Beyond this, I used the stories shared by the participants who completed each defining Q-sort as catalysts to further illuminate the meaning of each factor. 

Ultimately, labeling and describing the factors is an interpretative process. I undertook this process as a researcher experienced with qualitative methods---I primarily use phenomenological methods and thematic analyses in my work. In my role as an instructor, I have several years of experience teaching courses in UX for both undergraduate non-UX majors and master's UX majors, and I keep up with the news on the technology industry, particularly with respect to ethical issues. I have prior UX industry experience working in a marketing agency. All of these elements have informed my interpretative lens as a researcher on this project. 

\section{Findings}

Using factor analysis, I discerned five factors that cumulatively explain 75\% of the variance in participants' Q-sorts. (A minimum of 60\% is generally considered acceptable in exploratory factor analysis in the social sciences \citep{Hair2006}.) In the context of the research question of this study, these factors represent the workplace contexts that inhibit human-centered design in the experience of UX professionals. These factors are summarized in Table~\ref{table:summary} (the columns Speed and Vision are discussed in the ``Discussion'' section below).

Below, each factor is explained as interpreted through the context of the data from this study. Quotes from the participant stories collected with each defining Q-sort (Q-sorts which load more strongly on that factor than on other factors) are shared where such quotes are available (not all participants shared a story, and not all stories helpfully illuminate the factor in question).

\nohyphens{
\begin{table}[]
	\caption{Summary of the factors}
	\label{table:summary}
\begin{tabular}{@{} l p{1in} p{2in} l l l @{}}
\toprule
   & Factor~Name                     & Description                                                                                                                                        & Defining Sorts & Speed & Vision \\ \midrule
F1 & Single-Minded Arrogance         & A clear but flawed vision drives a project forward with little room for negotiation or adaptation                                                  & P3, P13, P1    & Fast  & Clear             \\
F2 & Competing Visions               & Disagreements and tensions among product stakeholders muddy the team's direction and ultimately manifest as antipathy toward UX                    & P8, P11, P9    & Slow  & Muddy             \\
F3 & Moving Fast and Breaking Things & Deferring UX activities to a later, unspecified date on the assumption that they are unnecessary                                                   & P12, P2        & Fast  & Clear             \\
F4 & Pragmatically Getting By        & The reality in which human-centeredness is seen as one lens competing with many others, in which many decisions are out of UX professionals' hands & P10            & Slow  & Muddy             \\
F5 & Sidestepping Responsibility     & Situations in which the ethical aspects inherent to design are dismissed or ignored                                                                & P7, P6         & Fast  & Clear             \\ \bottomrule
\end{tabular}
\end{table}
}

\subsection{Single-Minded Arrogance (Factor~1)}

The first factor, dubbed Single-Minded Arrogance, represents workplace contexts in which a clear but flawed vision drives a project forward with little room for negotiation or adaptation. I use the term ``arrogance'' here not in the sense of having inflated self-importance but in the sense of being incorrigible. P1 described this situation as ``\emph{anti-process},'' in which a manager appeals to ``\emph{common sense}'' rather than research findings or UX best practices. Similarly, P3 discussed the design process being rushed, without time dedicated to testing or iterating.

This factor explained 20\% of variance in the Q-sorts. The defining Q-sorts were those of P3, P13 and P1. Some of the distinguishing statements for this factor included: 

\begin{itemize}
	\item 20: \emph{Stakeholders not seeing the value of user-centered design} (Z-score 1.608)\footnote{A Z-score is a weighted average of the scores that similar participants gave to a statement. \citep{Zabala2018}}
	\item 2: \emph{Not doing research to understand user needs and values} (Z-score 1.369)
	\item 9: \emph{Lack of iteration} (Z-score 1.159)
	\item 28: \emph{Resistance to change} (Z-score 1.159)
\end{itemize}

\subsection{Competing Visions (Factor~2)}

The next factor, Competing Visions, represents disagreements and tensions among product stakeholders that muddy the waters of a team's direction and ultimately manifest, wittingly or not, as antipathy toward UX. Exemplifying this factor, P8 discussed how a knee-jerk response to fraud among users led to a redesign with bad UX, and P9 discussed how disagreements during the design stage of a project ended up minimizing the voice of UX on the team. 

This factor explained 19\% of the variance in the Q-sorts. The defining Q-sorts were those of P8, P11 and P9. Some of the distinguishing statements for this factor included: 

\begin{itemize}
	\item 2: \emph{Not doing research to understand user needs and values} (Z-score 1.718)
	\item 24: \emph{Politics between teams involved} (Z-score 1.492)
	\item 22: \emph{Sabotage, vendetta or retribution} (Z-score 1.25)
	\item 34: \emph{Lack of exploring ethical values} (Z-score 1.1)
\end{itemize}

\subsection{Moving Fast and Breaking Things (Factor~3)}

The third factor, Moving Fast and Breaking Things, takes its name from the motto coined by Mark Zuckerberg and used internally as a prime directive at Facebook until May 2014 \cite{Baer2014}. As the name suggests, this factor primarily represents deferring UX activities to a later, unspecified date. Regarding this factor, P2 described a situation where the founders of a startup ``\emph{make all the decisions based on their preferences rather than doing any research or validating anything.}''

This factor explained 12\% of the variance in the Q-sorts. The defining Q-sorts were those of P12 and P2. Some of the distinguishing statements for this factor included: 

\begin{itemize}
	\item 2: \emph{Not doing research to understand user needs and values} (Z-score 2.521)
	\item 7: \emph{Ignoring insights from user research} (Z-score 1.665)
	\item 3: \emph{Missing an important stakeholder group during research} (Z-score 1.329)
	\item 6: \emph{Planning to implement features later that are better aligned with user needs} (Z-score 1.329)
\end{itemize}

\subsection{Pragmatically Getting By (Factor~4)}

Pragmatically Getting By was the fourth factor discerned in this study. This factor represents the real-world exigencies of doing design, in which human-centeredness may be seen as one lens among others, and in which many decisions are out of UX professionals' hands. In this factor, design teams do the best they can with the sparse opportunities and resources they have. P10, for example, shared a case when the business structure was cemented before the design work began, and the design team had to use that (non--user-centered) structure as a constraint. In such cases, UX professionals are unable to ``keep up'' with the implications of decisions made by others. 

This factor explained 11\% of the variance in the Q-sorts. The defining Q-sort was that of P10. Some of the distinguishing statements for this factor included: 

\begin{itemize}
	\item 29: \emph{Not trying to create a user-centered product} (Z-score 1.716)
	\item 1: \emph{User needs conflict with each other} (Z-score 0.869)
	\item 2: \emph{Not doing research to understand user needs and values} (Z-score 0.869)
	\item 13: \emph{Compromises during technical implementation} (Z-score 0.869)
\end{itemize}

\subsection{Sidestepping Responsibility (Factor~5)}

The final factor, Sidestepping Responsibility, refers to situations in which the ethical aspects inherent to design are dismissed or ignored. To be sure, with complex projects and dynamic teams, it may be difficult to discern responsibility, decide who should take action, and so on. However, this factor points to situations in which responsibility is explicitly ignored. For example, P6 shared a story of colleagues emphasizing that the end product should primarily be aligned with company sales goals. ``\emph{In the end the product was what [they] wanted and not aligned to the user's needs.}.''

This factor explained 13\% of the variance in the Q-sorts. The defining Q-sorts were those of P7 and P6. Some of the distinguishing statements for this factor included: 

\begin{itemize}
	\item 29: \emph{Not trying to create a user-centered product} (Z-score 1.915)
	\item 11: \emph{Not enough time/resources spent in testing} (Z-score 1.275)
	\item 32: \emph{Using ethical toolkits incorrectly} (Z-score 1.275)
	\item 23: \emph{Lack of the necessary people taking responsibility} (Z-score 1.131)
\end{itemize}

\section{Discussion}
 
Across the fourteen subjectivities expressed in the Q-sorts, I identified five clusters representing workplace situations in which human-centered design is inhibited. 

These clusters are resonant with findings from previous literature, as reviewed in the ``Related Work'' section above. For example, the present findings reiterate that UX work is highly collaborative, particularly with teammates and other stakeholders outside UX \citep{Frishberg2020,Yang2018,Yang2020}, and that industry processes generally involve a stage of handoff \citep{Maudet2017,Yang2020,Feng2023} where the project leaves UX professionals' hands (``Pragmatically Getting By''); in such situations UX professionals must do what they can. The findings also reflect how sometimes non-UX stakeholders may be effectively hostile to UX because of multiple priorities (``Competing Visions'') and the centralization of power (``Single-Minded Arrogance'') \citep{Arledge2019,Gray2015,Patel2020}. Furthermore, since a novel (to HCI) methodology was used to arrive at these findings, this paper lends further credence to their trustworthiness.

At the same time, these five workplace situations, as clusters of barriers to human-centered design, can be useful for education, diagnostics and awareness. The literature on the realities of UX practice should be incorporated into the teaching of UX and human-centered design. As one idea to that end, building on the suggestion of a reviewer of this manuscript, the five factors from this study could be used as the basis to develop scenarios, case studies and company/project profiles. The descriptions of the factors offered in Table~\ref{table:summary} offer a starting point. Such scenarios or cases could be the subject of class discussions and other assignments (e.g.,~asking students to find real-world cases that match or contradict their assigned scenario). 

As presented above, each cluster has distinguishing characteristics and stands alone as an explanatory factor. Yet there are also certain shared similarities across these factors. In this discussion, I analyze these shared similarities to arrive at a deeper understanding of these factors and their relationships and to contribute to theory in ethical design, UX practice and HCI more broadly. 

\subsection{Two Dimensions Underpinning the Five Factors}

While analyzing the data from this study and interpreting the factors, I noticed that some of the factors did not seem mutually exclusive. For instance, a project could move forward by Moving Fast and Breaking Things (Factor~3) and Sidestepping Responsibility (Factor~5) with Single-Minded Arrogance (Factor~1). Indeed, as shown in Table~\ref{table:factorchars}, the factors correlate with each other to varying degrees; for example, Factor~1 correlates nearly 39\% with Factor~5 and 23\% with Factor~3, but only 11\% with Factor~4. Continuing the interpretive work inherent to Q methodology \citep[p.~14]{McKeown2013}, I sought to better understand how the clusters related to each other conceptually. 

As I reflected on the five factors, I first noticed that the dimension of speed clearly underlay some of the factors. For instance, Moving Fast and Breaking Things (Factor~3) explicitly suggests a high speed, while Pragmatically Getting By (Factor~4) and Competing Visions (Factor~2) imply (albeit more subtly) a lower speed. I also noticed that some factors connoted how clear the end state of the project was. Single-Minded Arrogance (Factor~1) suggests a very clear (if hubristic) vision of the end state, while others suggested a muddled end state (Competing Visions, Factor~2) or an emerging one (Pragmatically Getting By, Factor~4).

These ideas were also expressed in some of the stories shared by participants. For instance, P3's story implies quick movement and time pressure: ``\emph{My experience is mainly through freelance projects where the product owner wants to flesh out a complete product within a month from little to no research and user testing}.'' And P1's story suggests an unfortunately clear vision: ``\emph{The product owner was very anti-process and kept reiterating that everything was `common sense.'}{}''

\begin{figure}[h]
  \centering
  \includegraphics[width=0.5\linewidth]{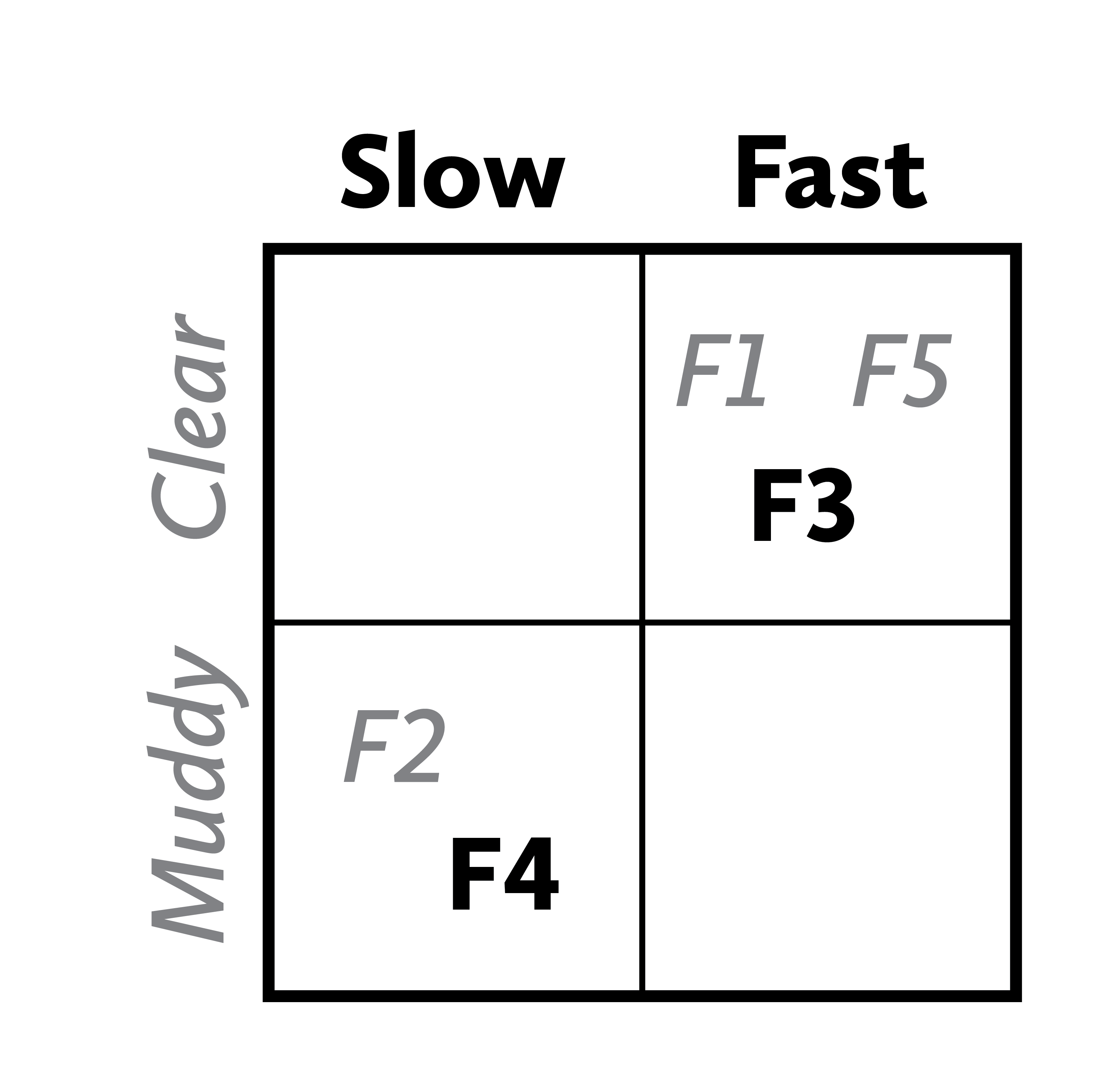}
  \caption{The five factors mapped in a two-by-two matrix according to the dimensions of speed (slow or fast) and clarity of vision (muddy or clear). The formatting indicates the dimension that most strongly defines a given factor.}
  \Description{A two-by-two matrix with the axes muddy--clear and slow--fast. In the box corresponding to muddy--slow, F2 and F4 are listed. In the box corresponding with clear--fast, F1, F3 and F5 are listed. The other two boxes are empty.}
  \label{fig:quadrants}
\end{figure}

I then explored how all the factors might map onto these two dimensions. I found that, in my interpretation, each factor primarily manifested one dimension and secondarily manifested the other. Thus I categorized the factors in a two-by-two matrix as shown in Figure~\ref{fig:quadrants}. Note that the labels Slow/Fast and Muddy/Clear are used for shorthand in this framework; in particular cases, an adjective such as ``circuitous'' may be more appropriate than ``slow,'' and ``emerging'' may be better than ``muddy.''

The factors Moving Fast and Breaking Things (Factor~3) and Pragmatically Getting By (Factor~4) were primarily defined by speed, with Factor~3 expressing high speed and Factor~4 low speed. The remaining factors were primarily defined by clarity of vision. Competing Visions (Factor~2) suggests a muddy overall vision, while Single-Minded Arrogance (Factor~1) and Sidestepping Responsibility (Factor~5) both represent clear visions. Interestingly, though Factors 1 and 5 both suggest clear vision, Factor~1 is defined by a positive vision (the direction the product is headed), while Factor~5 is defined by a negative vision (what certain stakeholders are trying to avoid).

The statistical analysis roughly supports this interpretation, but not perfectly. Looking at the factor score correlations (see Table~\ref{table:factorchars}), Factors~1 \& 5 and Factors~2 \& 4 are highly correlated (39\% and 35\%, respectively), as this two-dimensional analysis would suggest; and Factors~1 \& 4 and Factors~4 \& 5 are poorly correlated (11\% and 15\%), also as the two-dimensional analysis suggests. But at the same time, Factors~1 \& 2 and Factors~3 \& 4 are also highly correlated (33\% and 32\%), contrary to the two-dimensional analysis.

In the end, if the two-dimensional analysis were perfect, the factor extraction would have led to only two factors, not five. But there is sufficient resonance in the two-dimensional analysis to support further reflection. 

It is immediately evident that of the four logical possibilities, only two are represented by the data in this study. Since this study relates to situations in which human-centered design is thwarted, to me the absence of data in the other two quadrants suggested that those two quadrants could represent possibilities for doing human-centered design. I will return to this idea below. 

\begin{figure}[h]
  \centering
  \includegraphics[width=0.5\linewidth]{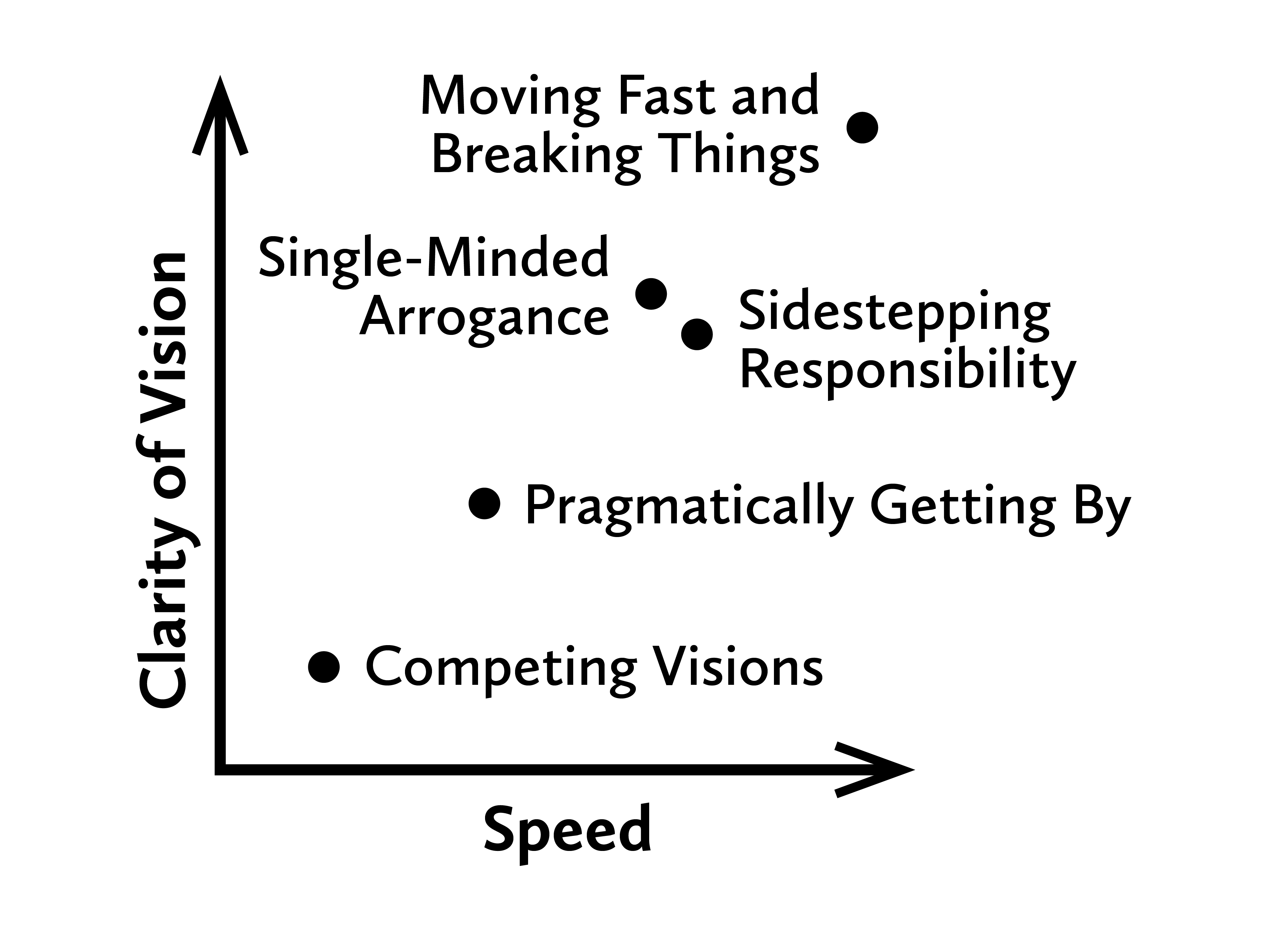}
  \caption{The five factors mapped on a coordinate plane with the axes of speed and clarity of vision.}
  \Description{A coordinate plane with the axes labeled clarity of vision and speed. The five factors are mapped in approximately a linear fashion. Starting from zero, the order is: pragmatically getting by, competing visions, then single-minded arrogance and sidestepping responsibility, and lastly moving fast and breaking things. Single-minded arrogance and sidestepping responsibility are approximately equal in position.}
  \label{fig:chart}
\end{figure}

In continuing my analysis, I did not feel the two-by-two matrix was sufficient to capture the full story. For example, as mentioned above, Factor~3 is primarily defined by speed, so its presence in the same box as Factor~1 and Factor~5 may be misleading. To address this, I attempted to chart these factors on a coordinate plane, again with the axes of speed and clarity of vision, as shown in Figure~\ref{fig:chart}.

Mapping the factors in this way forced me to judge, for instance, whether Factor~2 or 4 was ``slower,'' and which among Factors 1, 3 and 5 expressed the most clarity of vision. I judged Competing Visions (Factor~2) to be the slowest and muddiest because it suggests progress in minimal increments and with no shared vision. P8 and P9, who generated two of the defining Q-sorts for this factor, both shared stories of product failure because the team failed to cohere around a particular vision. In the case of P9, the product failed to launch entirely; in P8's story, the product launched but had ``\emph{a hostile and unclear user experience and a significantly smaller customer pool.}''

At the other end, I judged Moving Fast and Breaking Things (Factor~3) as representing the highest speed and clearest vision of all the factors. P2, who generated one of the defining sorts for this factor, expressed this well: ``\emph{I'm a consultant and I see this happening all the time. The most egregious cases are often in startups where the founders `are the user' and assume their experience is representative of all users and therefore they make all the decisions based on their preferences rather than doing any research or validating anything.}''

\subsection{Spaces of Opportunity for Human-Centeredness}

This study was about how things go wrong when it comes to human-centered design and UX ethics. But the findings offer some ideas for how to make things go right. Returning to the observation above that the results from this study only reflect a portion of the logical possibilities available in the two-by-two matrix, I would suggest that opportunities for enabling human-centered design may lie in the remaining spaces: moving fast with a muddy vision, and moving slowly with a clear vision.

Moving quickly with an unclear vision may sound like a recipe for disaster. But upon reflection, the notion is evocative of rapid prototyping, which is a well studied and now standard approach to human-centered design \citep{Culen2014,Mast2023}. As Buxton succinctly puts it, rapid prototyping is an approach for ``getting the design right and the right design'' \citep{Buxton2007}. The first part of that quote emphasizes that early in the design process, the team may not be sure what problem is being solved (i.e.,~the vision is muddy or emerging), let alone how to solve it. By progressing quickly through sketches, paper prototypes, Wizard-of-Oz prototypes and the attendant conversations, the team's vision gradually becomes clear. 

What this suggests for UX practitioners is that one way forward for further enabling human-centered design among product teams is---perhaps unintuitively---not to begin by advocating for the user's interests per se (which advocacy may fall on deaf ears), but to communicate the purpose and effectiveness of rapid prototyping. Championing and building such a design culture in general within an organization \citep[see][]{Justinmind2018,Kilian2015} may be a necessary foundational precursor to (or at the very least least a helpful support for) human-centered design and ultimately more ethical products. One path toward shifting the culture in this way is to begin by communicating the business case for prototyping (it saves money by not expending many resources toward what turn out to be dead ends), as well as leveraging prototyping techniques that elicit the most engagement from non-UX stakeholders \citep{RodriguezCalero2020}. Other specific tactics for promoting a culture of prototyping are offered by Buxton in \emph{Sketching User Experiences} \citep{Buxton2007} and, outside the design world, Ries in \emph{The Lean Startup} \citep{Ries2011}.

Next, moving slowly with a clear vision suggests caution and humility, not barreling full-speed ahead even when the destination is known. Consider how the situation of Single-Minded Arrogance (Factor~1) is characterized by resistance to change and a lack of iteration, and that of Moving Fast and Breaking Things (Factor~4) is characterized by short-changing the research phase. These observations suggest that a more human-centered way forward would be characterized by iteration, openness to change, and careful, valued research. 

Organizational psychologist Adam Grant counsels for just such an attitude and approach in his book \emph{Think Again: The Power of Knowing What You Don't Know} \citep{Grant2021}---and he does so from the perspective of business performance and profitability, which may make his arguments more palatable to corporations. Grant writes that in a world characterized by change (and technology projects are certainly not immune to change), effective organizations and teams must cultivate the skills of rethinking and unlearning. Put in the context of this paper: even if the vision seems clear, it's wise to be attentive to how it might need to evolve in response to outside forces. 

Throughout \emph{Think Again}, Grant reports on research results and shares case studies for building these skills individually, interpersonally and organizationally. Of note for the present discussion is his chapter on building cultures of learning at work, which discusses the essential tactics of cultivating psychological safety (see also \citep{Edmondson1999,Hsieh2023}) and creating incentives to reward rethinking. Infusing an organizational culture with the mindset of rethinking and relearning may, like building a design culture generally, create a workplace more conducive to human-centered design. 

Moving fast with a ``muddy'' vision and moving slowly with a clear vision present spaces of opportunity for putting human-centered design into practice. Without these foundations in place, it may not be possible to shoehorn conversations on ethical design and similar issues into an organization's culture. These findings are aligned with and extend those from prior studies, such as Chivukula et al. \citep{Chivukula2020} and Pillai et al.'s \citep{Pillai2022} discussions of how workplace culture influences the possibilities for success of UX efforts, and prior discussions of how organizational practices impinge on ethical action \citep{GrayChiv2019}. The findings from this study thus connect with other conversations across HCI (and CSCW in particular) on shifting organizational culture to facilitate ethical and human-centered design \citep{Rakova2021,Shneiderman2020}.

Many designers, researchers and other stakeholders today would like to see organizations take a more ethical approach to technology development. Yet guidance to that end has been limited. As reviewed above, the existing literature tends to be centered on the agency of UX professionals and includes theoretical contributions (e.g.,~mapping the complexities of acting ethically) and practical contributions (e.g.,~toolkits, card decks, activities). If these efforts are not yet resulting in widespread impact, one reason may be that examining ethics per se is already looking too far downstream (the water is blocked upstream, elsewhere in the organization). The findings from this study suggest a different approach: rallying first for shifts in organizational culture toward a culture of design and intellectual humility.  

To be sure, championing prototyping and humility will not remove budget constraints or incorrigible managers, nor will it straightforwardly shift an organization's culture. As one reviewer of this manuscript put it, if things move slowly, who has the power to change the pace? What if the UX designer's vision is clear, but stakeholders disagree? As \citet{Arledge2019} also writes, individual designers have limited power and must work within business and industry structures. For that reason, Arledge calls for designers to work to influence organizational and governmental policy. At present, designers and other UX advocates may practice ``soft resistance,'' leveraging the places they have agency to effect changes, however small or isolated \citep{Wong2021}. Based on the present study, the suggestion would be for the tactics of soft resistance to become strategic in the sense that they add up to something larger and are not ad-hoc efforts.

Larger, longer-term changes may have to come from above. Consider the example of Nest, mentioned in the introduction; in that case, the company was successful in operationalizing human-centered design perhaps mostly because the founder was already a proponent and practitioner. As well, Grant's suggestions for humility in \emph{Think Again} \citep{Grant2021} seem more practicable by managers and leaders than by individual contributors. 

One hope in this regard is that as the UX profession continues to mature, more managers and other stakeholders will see the value of UX.
As UX professionals progress in their careers and take on higher-level roles, they may be in a better position to make the case for the business value of human-centered design; the return on investment and long-term efficacy of human-centered design have been thoroughly documented \citep{Nielsen2016}. At the same time, in education, UX is beginning to be acknowledged as a key component of computing. At my institution, for instance, computing and software engineering majors take at least one course in human-centered design. Further on, though admittedly more pie-in-the-sky, new corporate formulations such as B Corporations,\footnote{See \url{https://www.bcorporation.net}} which explicitly value social and environmental contributions as well as profit-generation, may become more common in the tech industry. More generally, we can also hope and argue for corporations shifting from a short-term focus on quarterly performance to a longer view measured in years.

\section{Limitations and Future Work}

As an interpretivist study with a small group of participants, the findings from this research cannot be statistically generalized to broader populations. Rather, they can be analytically applied to other situations and contexts on a case-by-case basis. The results presented here should be taken as possible experiences rather than an exhaustive statement. While Q methodology is designed to produce trustworthy and inclusive results, this is in the end only one study. 

Further, the implementation of this study as an anonymous online survey introduced certain limitations. First, personal information on the participants was not collected and thus their identities and claims cannot be verified. However, given the recruitment strategy and the nature of this study, it is unlikely that any participants took part in bad faith. Next, the use of Qualtrics as the host for the Q-sort activity was regrettable in retrospect---the interface for card sorting left much to be desired. One participant commented that because of the way the boxes dynamically resize and how the cards are organized, it was difficult to complete the sorting task. While the gold standard method for conducting Q-sorts is through in-person sessions with physical cards, doing them online affords convenience and broader reach. For future Q-methodological studies online, other platforms should be considered. A list of platforms and considerations for online Q-methodological studies is offered by \citet{Meehan2022}; an additional platform not listed in their paper is Q Method Software (\url{https://qmethodsoftware.com}).

The study presented in this paper suggests a number of opportunities for future research. As Q methodology offers a toolkit for studying subjectivity (in this case that of UX professionals), participants' impressions are not necessarily correct or incorrect---they are perceptions and experiences. Further research could use other methods to study the institutional and organizational dynamics of UX work in practice to validate, challenge and extend the findings from this study, as well as to represent the perspectives of other stakeholders. In particular, case study research could prove valuable to this end. Interview research could also provide an opportunity for participants to clarify their views, go deeper on these topics, and raise issues that were not anticipated in the Q-set used in this study, which was based primarily on the prior literature and my own experience. Moreover, interviewing participants after completing the Q methodology analysis would provide an opportunity to collect additional testimony, stories and context from participants. Such further research could uncover deeper complexity to the issues discussed here; for instance, what seems fast or muddy to a UX professional may look different to those in the C-suite. 

Moreover, research with professionals outside the United States would be welcome, as it is possible that organizational dynamics differ by country, language or other cultural dimensions. Similarly, it may be useful to segment participants by industry (e.g.,~retail, finance, etc.) in future studies, as the dynamics under discussion may vary by industry.

Tangentially, this study also serves as an example of how Q methodology can be used in HCI, and the methodology could be leveraged to shed light on any number of other research questions at the intersection of human subjectivity and computing.

Next, following up on the discussion from this study, it would be worthwhile to create interventions for promoting a design culture and humility within organizations and to test whether these interventions serve to better enable human-centered design and ethical discourse within the organization. Doing so would not only impact teams and products, but it would also create stronger links between the literatures on human-centered design, design ethics and organizational change. 

Finally, this study suggests the creation of curricular and other educational materials for UX students that better prepare them for the realities of the workplace. As discussed by \citet{Watkins2020}, for example, current UX and design education may serve to help students develop idealistic, unimplementable philosophies; stronger education on the realities of the discipline may encourage students to find ways to bridge the gap. The idea of creating scenarios from the findings from this study was mentioned at the beginning of the ``Discussion'' section. These scenarios could also be used in research with tech sector professionals as a technique for elicitation in interviews.

\section{Conclusion}

This paper has sought to shed light on the barriers to human-centered design in practice, through a Q-methodological study of UX professionals' experiences on teams that deviated from human-centered design best practices. With factor analysis, these experiences were divided into five clusters: Single-Minded Arrogance, Competing Visions, Moving Fast and Breaking Things, Pragmatically Getting By, and Sidestepping Responsibility. Beneath these five factors, two deeper dimensions were discerned: speed and clarity of vision. In the experiences of the participants in this study, non--human-centered design resulted from moving slow with an unclear vision and from moving fast with a clear vision. The remaining two logical possibilities---moving fast with an emerging vision and moving slowly with a clear vision---suggest ways to enable human-centered design: first, through cultivating a design culture that respects and understands the business case for rapid prototyping and testing; and second, through building a culture of intellectual humility that values rethinking and relearning in our fast-changing world. 

\bibliographystyle{ACM-Reference-Format}
\bibliography{../../refs}

\received{January 2024}
\received[revised]{July 2024}
\received[accepted]{October 2024}

\end{document}